\documentclass[a4paper,11pt]{article}
\pdfoutput=1 

\usepackage{jcappub}

\usepackage{multirow}
\usepackage{xcolor}
\usepackage{graphicx}
\usepackage{stmaryrd}
\SetSymbolFont{stmry}{bold}{U}{stmry}{m}{n}
\usepackage{comment}
\usepackage{enumitem}

\usepackage{dcolumn}
\usepackage{bm}

\newcommand{\cmnt}[1]{}

\newcommand{\prd}{Phys. Rev. D}
\newcommand{\prl}{Phys. Rev. Lett.}

\newcommand{\aap}{Astron. Astrophys.}

\newcommand{\apj}{Astrophys. J.}

\newcommand{\apjl}{Astrophys. J. Lett.}
\newcommand{\mnras}{Mon. Not. R. Astron. Soc.}

\newcommand{\jcap}{J. Cosmol. Astropart. Phys.}

\begin{document}
\title{Effects of Primordial Black Holes on IGM History}
\date{\today}

\author[a,b,c]{Emily Koivu,}
\author[c,e,f]{Nickolay Y.\ Gnedin,}
\author[a,b,d]{Christopher M.\ Hirata}
\affiliation[a]{Center for Cosmology and AstroParticle Physics, The Ohio State University, 191 West Woodruff Avenue, Columbus, OH 43210, USA}
\affiliation[b]{Department of Physics, The Ohio State University, 191 West Woodruff Avenue, Columbus, OH 43210, USA}
\affiliation[c]{Theory Division, Fermi National Accelerator Laboratory, Batavia, IL 60510, USA}
\affiliation[d]{Department of Astronomy, The Ohio State University, 140 West 18th Avenue, Columbus, Ohio, 43210, USA}
\affiliation[e]{Kavli Institute for Cosmological Physics, The University of Chicago, Chicago, IL 60637, USA}
\affiliation[f]{Department of Astronomy \& Astrophysics, The University of Chicago, Chicago, IL 60637, USA}
\emailAdd{koivu.1@osu.edu}

\abstract{Currently the asteroid mass window (mass $\sim 10^{17}- 10^{21}$ grams) remains unconstrained for Primordial Black Holes (PBHs) to make up all of the dark matter content of the universe. Given these PBHs have very small masses, their Hawking temperature can be up to hundreds of keV. This study investigates the potential impacts of PBH Hawking radiation on the intergalactic medium from $z\sim 800-25$, namely studying the ionization history, kinetic gas temperature, and ultimately the 21 cm signature. We find that for masses on the low edge of the asteroid mass window, there are up two orders of magnitude increases in the ionization fraction and kinetic gas temperature by redshift 25, and the 21 cm spin temperature can differ from non-PBH cosmology by factors of a few. This analysis results in maximum differential brightness temperatures of +17 mK for our lightest PBH masses of $2.12\times 10^{16}$g. We also show maximal $53$ mK discrepancies in differential brightness temperatures between our PBH and non-PBH cosmologies for our lightest PBH mass, while our heaviest PBH mass of $1.65 \times 10^{17}$g shows only $0.5$ mK variations. We find the Hawking-radiated electrons and positrons are instrumental in driving these IGM modifications. This study shows the necessity for a rigorous treatment of Hawking radiation in PBH cosmological observables from the dark ages through cosmic dawn.}

\maketitle

\section{Introduction}

The fundamental nature of dark matter is one of the greatest research questions in science today. Despite being an active research question for almost a century, we still have have a wide variety of candidates for dark matter that have yet to be ruled out. Present in that candidate collection is Primordial Black Holes --- theorized black holes created in the early universe from density perturbations rather than typical stellar collapse \cite{Villanueva-Domingo:2021spv,PhysRevD.94.083504,BIRD2023101231}. Primordial Black holes (PBHs) are an interesting candidate because while their formation requires new physics in the early Universe, they do not require any long-lived new particles beyond the Standard Model.

There are constraints on PBHs in most mass ranges from a variety of observables \cite{2021RPPh...84k6902C}. For large masses ($>100M_\odot$), the Poisson fluctuations in the number of PBHs act as an additional source of perturbations on small scales \cite{1983ApJ...268....1C}, which has led to constraints from the Lyman-$\alpha$ forest \cite{2003ApJ...594L..71A, 2019PhRvL.123g1102M}. The stellar-to-planetary mass range has been probed with gravitational microlensing, ranging from magnification scatter in Type Ia supernovae at large masses \cite{2018PhRvL.121n1101Z}, through stellar microlensing searches \cite{2007A&A...469..387T, 2019PhRvD..99h3503N}, and recently through searches for short-duration (minutes) microlensing events \cite{2014ApJ...786..158G, 2019NatAs...3..524N, 2020PhRvD.101f3005S}. Below masses of $\sim 10^{23}$ g, finite source size effects make the microlensing route increasingly difficult: for fixed observational sensitivity, the rate of detectable events scales as $\propto M_{\rm PBH}^4$ \cite{Montero-Camacho_2019}; and at $< 10^{22}$ g diffractive effects \cite{1986ApJ...307...30D} wash out the optical microlensing signal. At very low masses ($\lesssim 10^{17}$ g), the Hawking radiation (volume emissivity $\propto 1/M_{\rm PBH}^3$) becomes a useful constraint: this includes the isotropic gamma ray background \cite{2010PhRvD..81j4019C}, as well as constraints from the local lepton spectrum \cite{2019PhRvL.122d1104B} and the Galactic 511 keV line \cite{2019PhRvL.123y1101L, 2019PhRvL.123y1102D, Ray1} (which will be probed further with the upcoming launch of the Compton Spectral Imager \cite{2023JCAP...02..006C}). This leaves an ``asteroid-mass window'' in the range of $\approx 10^{17}-10^{22}$\ g where the fraction of dark matter (DM) that could be in the form of PBHs is unconstrained.

The lightest end of this regime is bounded by Hawking radiation constraints. In order to search for these candidates, we are aided by the fact that Hawking radiation from black holes has a temperature which is inversely proportional to the mass: $T_H \approx (10^{16}~{\rm g}/M_{\rm PBH})$ MeV \cite{Hawking}. Therefore, these low mass primordial black holes can create gamma rays which are potentially observable in direct detection experiments. The PBH photon Hawking radiation spectrum also contains an extended low energy X-ray tail which is created by the $\mathcal{O}(\alpha_{\rm EM})$ inner-bremsstrahlung processes, i.e. ones where the photons are not directly sourced from the black hole but are allowed first-order interactions with the electrons/positrons also produced from the black hole \cite{Page:2007yr, Coogan:2020tuf}. This extended spectrum means that we can also look for potential effects of these asteroid mass PBHs in different cosmological observables. 

One of the most direct ways we can investigate PBHs as dark matter is looking for the effect of their Hawking radiation on the intergalactic medium (IGM) throughout history. In particular, PBHs could lead to a non-standard IGM evolution in the Dark Ages (before astrophysical sources of radiation such as stars or active galactic nuclei were present).
This potential signature therefore necessitates a detailed account of all of the potential processes through which PBH radiation might impact the IGM history. These photon-baryon processes consist of ways to ionize, heat, or excite the medium. As such, we aim to investigate cosmological observables with and without PBHs in order to determine whether there are potentially measurable differences in these two cosmological scenarios. This paper will not be the first of its kind; in fact there are a fair number of investigations of IGM signals of PBHs \cite{21cmfromlowmassPBH, EvaporatingPBHon21cmSignal, LowMassPBHEvapAng21CMForecast,Mittal_2022,khan2025strongerconstraintsprimordialblack}. The novelty in this work is due to the rigorous treatment of Hawking radiation from a first-principles QFT on curved spacetime calculation in \cite{HR1, HR2} not considered in previous numerical Hawking radiation codes, as well as the careful investigation of Compton scattering and the redshift-dependent energy deposition mechanisms for Hawking-radiated particles. 

In this paper, we will work through semi-analytic calculations for the imprints of PBH dark matter. We will discuss our plan for the calculation in Section~\ref{sec: background}.
We will then calculate the potential effects of PBHs on the ionization history in Section~\ref{sec: ionization}. Then, we will turn to impact of PBHs on the thermal history of the IGM in Section~\ref{sec: thermal}. Finally, we will look at how PBH radiation producing excitations of the IGM would manifest in the 21 cm global signal in Section~\ref{sec: 21cm}. This paper will conclude with a discussion the ramifications of all signatures for future surveys, as well as potential areas for further investigation in Section~\ref{sec: discussion}.

\section{Plan of the calculation}
\label{sec: background}

\begin{figure}
    \centering
    \includegraphics[width=\linewidth]{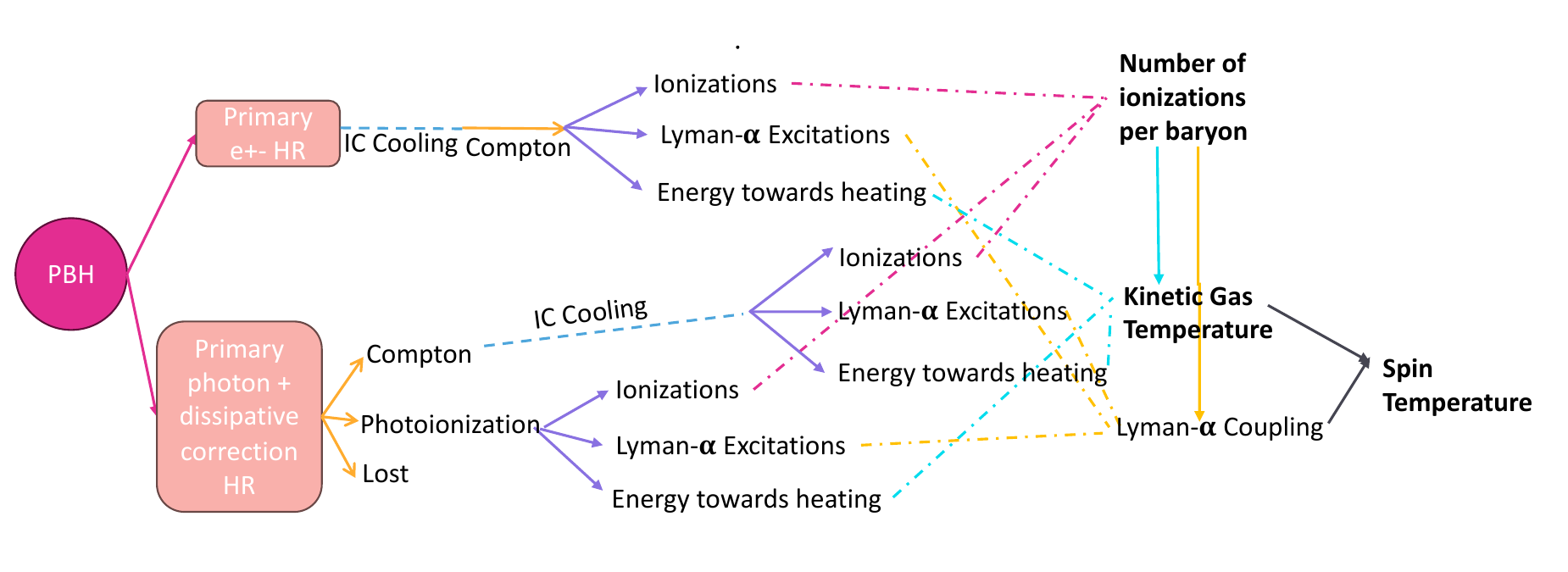}
    \caption{This flowchart describes the basic numerical engine that we utilize in this project. First, we generate our photon and fermion Hawking radiation spectra. We then compute the fraction of radiated photons then can photoionize and Compton scatter, and the rest are lost. If the photons Compton scatter, they must first undergo inverse Compton cooling. The fermions all inverse Compton cool and Compton scatter. From here, the radiation can contribute to ionizations, Lyman-$\alpha$ excitations, or heating the IGM. Based on which branching path the radiation falls under, we then use that computation to determine the ionization history. Next, we use this information to compute the kinetic gas temperature. The ionization history and gas temperature then also contribute to the Lyman-$\alpha$ coupling, and all of this information is used to compute the spin temperature.}
    \label{fig:flowchart}
\end{figure}

Our computation starts with the spectrum of photons ($\gamma$) and charged leptons ($e^\pm$) emitted by the PBHs, then proceeds through energy deposition mechanisms, and finally arrives at the ionization and temperature evolution of the IGM.
An outline of the general numerical structure can be found in Figure~\ref{fig:flowchart}. 

\subsection{Hawking radiation}

We use the results of the Hawking radiation spectrum derived analytically in \cite{HR1} and numerically in \cite{HR2}. This set of papers takes a Schwarzschild black hold background and performs a canonical quantization of QED fields, then calculates the first order correction to the dissipative Hawking radiation photon spectrum using perturbation theory.  In order to properly sample all energy ranges allowed in the HR spectrum, we will use a fitted model to interpolate between the energies not directly calculated in \cite{HR2}. Appendix~\ref{ap:HR} contains modeling information where we include the order $\mathcal{O}(1)$ photon and fermion spectrum, as well as the order $\mathcal{O}(\alpha_{\rm EM})$ correction to the photon spectrum dominated by inner bremsstrahlung processes.

The following analysis assumes that all dark matter is in the form of primordial black holes with a single specified mass (see Ref.~\cite{2018JCAP...04..007L} for a discussion of how single-mass constraints map onto scenarios with broader mass distributions). We take masses of 1, 2, 4, and $8\times 10^{21}m_{\rm Planck}$ (as used in the tabulated numerical results \cite{HR2}). We will hereafter refer to these masses as M1 = $2.12\times 10^{16}$ g, M2 = $4.24\times 10^{16} $g, M3 = $8.48\times 10^{16} $g, and M4 = $1.648\times 10^{17}$g, respectively.

\subsection{Energy deposition mechanisms}
\label{sec: interactions}

The spectra of the PBH Hawking radiation shows us that the energy range of interest scales from $\sim 10$ eV (in the inner bremsstrahlung tail, but where the photons have enough energy to excite atoms) to $\sim1$ MeV (the peak of the ``direct'' blackbody spectrum for the lower-mass PBHs). As such, the dominant interaction processes between the PBH-produced photons and the IGM will be photoionization at lower energy and Compton scattering at higher energy, though in principle both processes are included for all energy ranges. The photoionization equations for Hydrogen and Helium used in this work are taken from \cite{Summefeld,PIcrosssec}, and explicitly given in Appendix~\ref{ap:interactions}. The Compton scattering cross section is given by the Klein-Nishina equation, again explicitly shown in Appendix~\ref{ap:interactions}.

The energy loss of electrons and positrons in the IGM is a competition between collisional losses (which inject energy into the gas)\cite{Rohrlich,uehling} and inverse Compton losses (which inject energy into a spectral distortion in the CMB) \cite{ICCR}. The collisional losses are more important at low energy, whereas the inverse Compton effect is more important at high energy. (See, e.g., Fig.~2 of Ref.~\cite{Yang_2024} for an illustration.)
At all redshifts and energies of interest in this paper, at least one of the two energy loss mechanisms is efficient, so electrons and positrons will lose their energy. But any electron or positron that is injected above this ``crossover energy'' will efficiently up-scatter the CMB and lead to infrared photons, losing energy until it reaches the energy where collisional energy loss dominates. Then standard physics of energy deposition by charged particles in a gas kicks in. Thus, any electron in Compton channels who has energy above this crossover energy will be reassigned the energy of the crossover energy before the calculation continues. 

Photons are different, since the Universe can be optically thin or optically thick depending on energy and redshift. We define the optical depth as 
\begin{equation}
    \tau = \left[n_{\rm H} (\sigma_{\rm H,PI}+ \sigma_{\rm H,compt}) + n_{\rm He}(\sigma_{\rm HeI, PI}+\sigma_{\rm He, compt})\right]r_{\rm Hubble},
\end{equation}
where $n_{\rm H}$ and $n_{\rm He}$ are the physical number densities of the two elements; $\sigma_{\rm H,PI}$ and $\sigma_{\rm He,PI}$ are the photoionization cross sections; $\sigma_{\rm H,compt}$ and $\sigma_{\rm He,compt}$ are the Compton scattering cross sections; and $r_{\rm Hibble}$ is the physical Hubble length.
which will become useful for the following calculations. We use the cosmological parameters from Planck 2018 results \cite{Planck}.

\section{Ionization history}
\label{sec: ionization}

\subsection{Implementation}

In this study, we will first investigate the potential impacts PBH Hawking radiation might have on the ionization history. The PBH could produce photons which ionize the IGM in a number of ways: directly, through secondary ionizations --- where initial photons from the PBH might be scattered or absorbed in the IGM and produce secondary electrons --- or through tertiary events, where these released electrons then produce further collisional ionizations before thermalizing. We also consider collisional ionization due to the Hawking-radiated fermions. 

To compare the ratio of baryons to ionizations potential, we must first find the rate of ionizations produced by the Hawking radiated photons, given by 
\begin{eqnarray}
    \dot N_{ion,\gamma}(z) &=& \int (1-e^{-\tau(z)})\frac{dN_{\gamma}}{d\omega dt} \Bigg(1+  \nonumber \\
    && \left. \sum_{i=\rm H,HeI}\left[\frac{\hbar \omega -I_i}{I_i}p_{i,\rm ion}(\omega) g_{i,\rm PI} + \frac{h_{\rm IC}(\hbar\omega)}{I_i}p_{i,\rm ion}(h_{\rm IC}(\hbar\omega))g_{i,\rm compt}\right]\right) d\omega.~~~
    \label{eq:ndot_phot}
\end{eqnarray}
Here, $dN_\gamma/d\omega\,dt$ is the photon emission spectrum.  The ``1'' in this expression represents the electron ejected by photoionization or Compton scattering. Additional terms are included from subsequent ionization processes. We introduce
\begin{equation}
g_{i,a} = \frac{n_i  \sigma_{i,\rm a}}{\sum_{j=H,He}~n_j (\sigma_{j,\rm PI}+ \sigma_{j,\rm compt})},
\end{equation}
which is the branching probability of undergoing process $a$ (photoionization or Compton scattering) on target atom $i$ (H or He).

For photoionization (which is dominant at low energies), the resulting electron produces a mean number of tertiary ionizations given by 
\begin{equation}
    p_{i\rm ion}(\hbar \omega) = f_{1,i}(\hbar \omega) y_{1,i} + f_{2,i}(\hbar \omega) y_{2,i},
\end{equation}
where $f_{k,i}$ and $y_{k,i}$ are as defined by Ref.~\cite{RicottiGnedinShull}.

For Compton scattering (which is dominant at high energies), the resulting electron might cool first.
Here $h_{\rm IC}(\hbar \omega)$ is the energy of the secondary electron from a Compton scattering event after inverse Compton cooling. We include a probability weighting given by $(1-e^{-\tau(z)})$, showing the likelihood of these ionizing photons to encounter a gas particle at a given redshift. 

We also consider the effect on the ionization history from the Hawking radiated fermions, determined by 

\begin{equation}
    \dot N_{ion,e^{\pm}}(z) = \int \frac{dN_{e^{\pm}}}{dh\, dt} \sum_{i=H,He}\frac{h_{\rm IC}(h) - m_e c^2}{I_i} p_{,\rm ion}\bigl(h_{\rm IC}(h) - m_e c^2\bigr) \,dh, 
    \label{eq:ndot_e}
\end{equation}
where $h$ represents the electron energy. Again, $h_{\rm IC}$ represents the energy after the electron has inverse Compton cooled to the crossover energy (if applicable).

The sum of Eqs.~(\ref{eq:ndot_phot}) and  (\ref{eq:ndot_e}) is integrated numerically for photon and fermion energies from $13.6 $eV (the ionization energy of hydrogen) to $48.6$ MeV (in the far Wien tail of the Hawking radiation for our lowest PBH mass). This wide energy range allows us to sample the entire Hawking radiation spectrum spectrum of the PBH and therefore account for all potential interaction processes. 

The above calculation is due to a single PBH, so in order to understand the global impact of PBH dark matter, we compute the photon number density from the PBH that will interact with the IGM and compare that to the abundance of gas by 

\begin{equation}
    \dot{x}_{\rm ion, PBH} = \frac{\dot{n}_{ion}}{n_{gas}} (z) = (\dot{N}_{ion,\gamma}(z)+\dot{N}_{ion,e^{\pm} }(z) )\frac{\rho_{DM}}{m_{DM}} \frac{1.2m_{H}}{\rho_{gas}} 
    \label{eq:ngammanb}
\end{equation}

where $m_H = 1.67\times10^{-24}$g, and the 1.2 weighting factor is to account for Helium in the IGM. 

Based on an initial calculation, we find the ionization fraction to be non-trivially impacted by PBH DM of some of our masses; therefore, we must proceed with a chronological evolution of the IGM based on ionization fraction modifications to compute the ionization fraction, heating, and excitation history. As such, we use a simple forward Euler scheme to track the ionization history:  
\begin{equation}
    x_{\rm ion}(z+\Delta z) = x_{\rm ion}(z) - \frac{(\dot{x}_{\rm ion, PBH}
    -\alpha_{\rm A} x_{\rm ion}^2 n_H)\Delta z}{(1+z)H(z)}, 
\end{equation}
where $\Delta z<0$ (evolving forward in time) and $\alpha_{\rm A}$ is the Case A Hydrogen recombination coefficient \cite{recomb}. This is initialized at $z=800$, which is after CMB-driven ionizations from the excited levels of hydrogen cease to be important \cite{1999ApJ...523L...1S, 2011PhRvD..83d3513A}.

\subsection{Results}

\begin{figure}
    \centering
    \includegraphics[width=\linewidth]{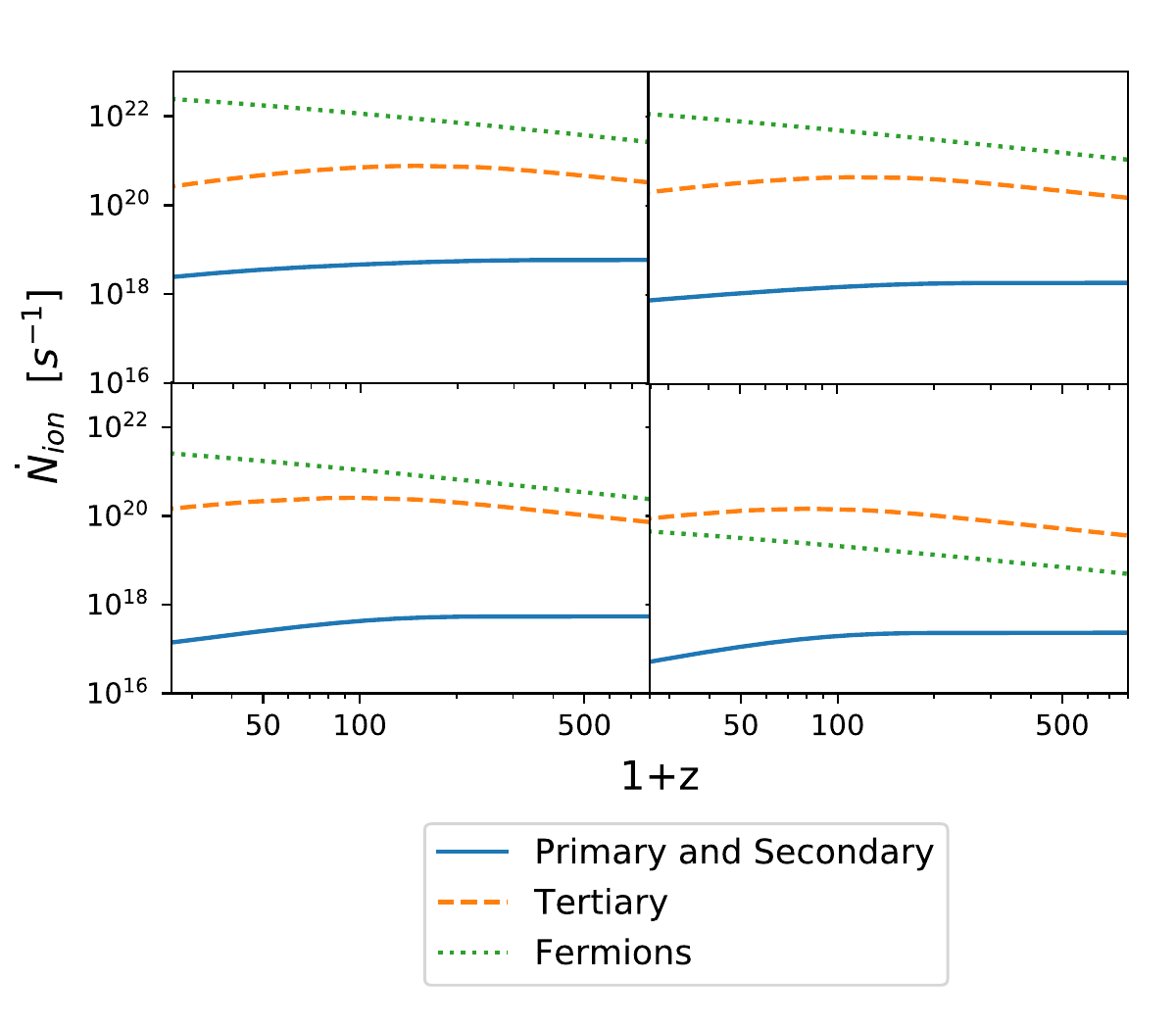}
    \caption{Rate of ionization produced by each ionization mechanism over time. Each panel corresponds do a different black hole mass (M1: top left, M2: top right, M3: bottom left, M4: bottom right). Here we see a slight decrease in all channels as mass increases. The most dominant ionization channel for smallest masses is through the Hawking radiated fermions, followed by the ionizations caused by `tertiary' processes of Hawking radiated photons. The fermion contribution has the largest suppression as mass increases. }
    \label{fig:Ndotchannels}
\end{figure}

\begin{figure}
    \centering
    \includegraphics[width=0.7 \linewidth]{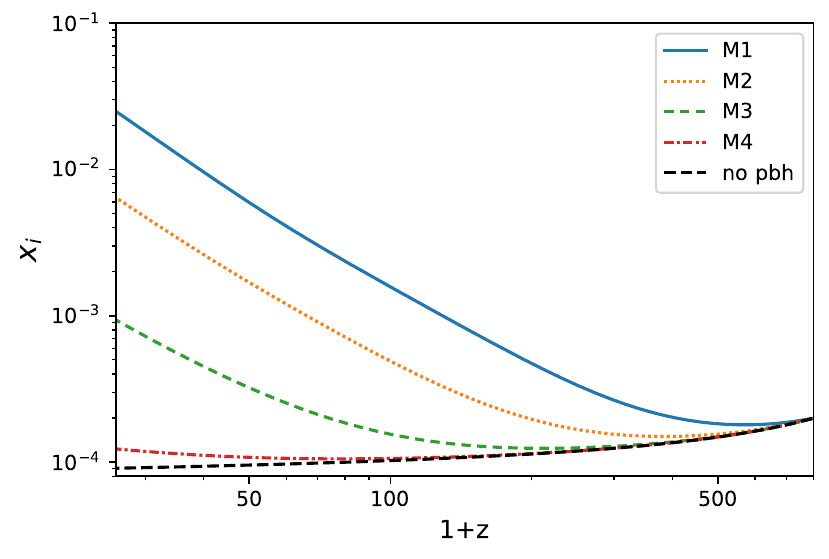}
    \caption{Ionization history for PBH dark matter masses (M1 solid blue, M2 dotted orange, M3 dashed green, M4 dotdashed red) as well as a fiducial no PBH cosmological scenario (in tightly-dashed black). For the lightest black hole mass, we see over a two order of magnitude difference in the ionization history relative to a standard cosmology. The deviations diminish as we increase mass, and fully disappear at the largest black hole mass we test. M1 shows the largest changes at lowest redshifts, reaching factors over $10^2$, while M4 deviates maximally by a factors on the order of $10^{-2}$.}
    \label{fig:NgammaoverNb}
\end{figure}

In order to first develop intuition on the sources of ionization, the importance of each of different ionization channel to the overall history can be found in Figure~\ref{fig:Ndotchannels}. Here we see that the ionizations due to Hawking radiated fermions are the dominant factor for most masses, followed by tertiary ionizations from Hawking radiated photons, and the total of primary and secondary are effectively negligible. Specifically, we see the ionization rates of M1 are approximately $10^{22}$ ionizations per second from fermions, a next-leading contribution of  $10^{20}$ ionizations per second from the tertiary photon products, and $10^{18}$ ionizations per second from the primary and secondary photon ionizations. As the PBH mass increases, we see a drastic suppression in the fermion spectrum of three orders of magnitude from the lighted to heaviest PBHs. This effect can be attributed to several factors but largely can be understood by the Hawking radiation spectrum. The peak of the fermion spectrum is at smaller energies for higher masses, which is approaching the rest energy of the electron, leading to only a small energy window of non-trivial Hawking radiation.

The overall evolution of $x_{\rm ion,PBH}$ from each of the masses can be seen in figure \ref{fig:NgammaoverNb}, as well as the relative difference of cases with PBH dark matter relative to without. We see a steady increase in this ionization fraction largely due to the Hawking radiated fermions for the M1, M2, and M3, with ionization fractions  reaching a maximum of $\sim 0.02$ for M1 by $z=25$. The ionization history steadily climbs during early times in our cosmic history where it would otherwise be difficult to modify the ionization fraction. We compute this effect through redshift $z=25$, which is where we expect stellar ionization channels to become important, and which we neglect in the scope of this paper. The most rapid change in the ionization history occurs for the lowest masses, but all masses but the highest exhibit a sizable change in the ionization fraction at $z= 100$. The heaviest mass does reach percent level differences in the ionization fraction by $z= 25$ in this analysis. 

Because of this nontrivial ionization history modification, in the following calculations, we proceed with a time ordered numerical evolution. At each timestep, we compute the ionization fraction by taking the previous ionization fraction and modifying it by Eqs.~(\ref{eq:ndot_phot}) and (\ref{eq:ndot_e}). We then compute all further parameters at the timestep using this new ionization fraction.

\section{Thermal History}
\label{sec: thermal}
\subsection{Implementation}

We now turn our attention to the thermal history of the IGM and the impacts of PBHs on the gas temperature. The PBH Hawking radiation spectrum spans a wide range of energies, and thus we must account for lower energy process such as photoionization and higher energy Compton scattering as both could dump heat into the surrounding medium. Additionally, while the IGM is primarily consistent of Hydrogen, we also include Helium in our calculation for completeness and because there are regions of energy where Helium cross sections are much larger than Hydrogen.  

To determine the energy injection rate from PBHs that contributes to heating the IGM, we will calculate the heating contributions separately by IGM species and interaction process. This requires us to compute the branching ratios of each of these heating mechanisms. For a species $i$, the heating energy injection rate from the Hawking radiation photons of a single PBH is given by 
\begin{eqnarray}
    \dot E_{i,\rm PI} &=& \int_{0}^{\infty} \frac{dN_{\gamma}}{d\omega dt} (1-e^{-\tau(z)}) (\hbar\omega-E_{0,i}) p_{i,\rm heat} g_{i,\rm PI}\, d\omega ~~~{\rm and}\nonumber \\ 
     \dot E_{i,\rm compt} &=& \int_{0}^{\infty} \frac{dN_{\gamma}}{d\omega dt} (1-e^{-\tau(z)}) L_{\rm compt} p_{i,\rm heat} g_{i,\rm compt} \,d\omega
\end{eqnarray}
where $E_{0,i}$ is the ground ionizing energy of a species $i$,  $p_{i,\rm heat}$ is the heating efficiency factor for each particle species \cite{RicottiGnedinShull} given by 
\begin{equation}
    p_{i,\rm heat} = 1-f_{1,i} y_{1,\rm heat} + f_{2,i} y_{2,\rm heat},
\end{equation}
and
\begin{equation}
    L_{\rm compt} = h_{\rm IC}(\hbar\omega) \frac{h_{\rm IC}(\hbar \omega) /m_e c^2}{\left[1+1.07\left(h_{\rm IC}(\hbar \omega)/ m_e c^2\right)^{3/5}\right]^{5/3}}
\end{equation}
is the angle-averaged relativistic Compton scattering resultant energy (derived from \cite{Compton}). By adding all contributions by IGM species and interaction mechanisms, we arrive at the energy injection from hawking radiated photons that contributes to heating. 

We must also compute the heating contribution from Hawking radiated fermions, given by 
\begin{equation}
      \dot E_{i,\rm compt} = \int_{0}^{\infty} \frac{dN_{e^{\pm}}}{dh dt}  h_{\rm IC}(h) p_{i,\rm heat}\,dh .
\end{equation}

The sum of these contributions is the energy injection from only one PBH of a given mass, so the total volumetric energy injection rate (in erg/cm$^3$/s) is given by
\begin{equation}
    \dot U_{\rm PBHs} = \frac{\bar \rho_{\rm DM}}{m_{\rm DM}} \dot E_{total}.
\end{equation}
The observable of interest is actually the impact of PBHs on the gas temperature; since we are focused on a period in cosmic history when the IGM was a monatomic gas, we write 
\begin{equation}
    \dot T_{PBHs}(z) = \frac{2}{3 k_{b}} \frac{\mu}{\bar \rho_{\rm gas}} \dot U_{\rm PBHs},
\end{equation}
with $\mu$ being the mean molecular weight and $\bar \rho_{\rm gas}$ being the mean density of the gas.

Given this additional heating source by PBHs, we are interested in assessing the overall gas temperature. This is given by 
\begin{equation}
    \frac{dT_{\rm K}}{dt} = -2HT_{\rm k} + \frac{x_{\rm ion}}{1+x_{\rm ion}} \frac{8\sigma_{\rm T}aT_{\rm CMB}^{4}}{3m_e c^2}\left(T_{\rm CMB}-T_{\rm K}\right) + \dot T_{\rm PBHs},
\end{equation}
where $x_{\rm ion}$ is the ionization history calculated in Sec.~\ref{sec: ionization}. We initialize at high redshift when the matter temperature is tightly coupled to the CMB and thus we can take $T_{\rm K,init} = T_{\rm CMB,init} = 2.72(1+z_{\rm init})\,$K.

The differential equation above is linear in $T_{\rm k}$, but it is stiff. To avoid numerical instabilities, we use an implicit (backward Euler) type solver:
\begin{equation}
    T_{\rm K} (z+\Delta z) = \frac{T_{\rm K}(z) + \Delta z \left(\frac{x_{\rm ion}}{1+x_{\rm ion}}\frac{8\sigma_{\rm T}aT_{\rm CMB}^{5}}{3m_e c^2} \frac{1}{(1+z)H(z)}\right)}{1 - \Delta z\left(\frac{2}{1+z} + \frac{x_{\rm ion}}{1+x_{\rm ion}}\frac{8\sigma_{\rm T}aT_{\rm CMB}^{4}}{3m_e c^2}\frac{1}{(1+z)H(z)} \right) },
\end{equation}
where again $\Delta z<0$.

\subsection{Results}

\begin{figure}
    \centering
    \includegraphics[width=0.7\linewidth]{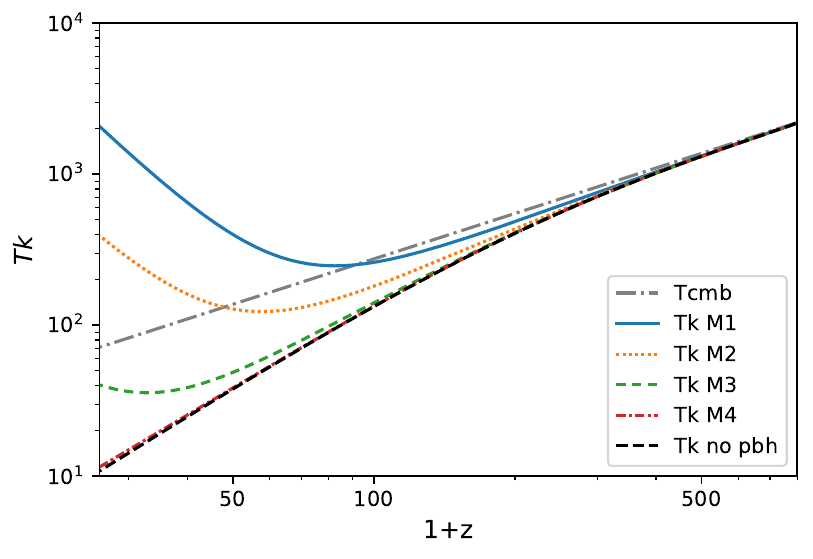}
    \caption{Thermal History [in K] of gas with PBH dark matter (M1 solid blue, M2 dotted orange, M3 dashed green, M4 dotdashed red) and without (in black tightly-dashed line) the additional heating of PBHs. Also plotted is the CMB temperature in dash-dotted gray. We see that at the earliest times, there is very little deviation from the no-PBH kinetic gas temperature while it is tightly coupled to the CMB temperature. As we move to later times, we see the kinetic gas temperature for the lightest three PBH masses all begin to deviate from the no-PBH cosmology. The two lightest masses even overcome the CMB temperature at redshifts of $z\sim 80$ and $z\sim 45$. The most extreme deviation of IGM kinetic gas temperature is with M1, seeing over a two order of magnitude deviation from the no-PBH example. We again see factors well over 100 difference between the no-pbh and M1 cosmology, and see percent level changes in the M4 scenario at latest times.}
    \label{fig:Tk}
\end{figure}

In Figure~\ref{fig:Tk}, we see a several order of magnitude difference appear in the thermal history between the case with PBHs and without for again the smallest black hole masses. Interestingly, we see the lightest two masses we investigate in this study create IGM heating so efficiently that, rather than relax to the typical $T(z) \sim (1+z)^2$ behavior, this early heating source is able to surpass CMB temperatures around $z\sim 100$ and $z\sim 45$, respectively. While the kinetic gas temperature from M3 does not surpass the CMB temperature by our model cutoff of $z= 25$, there still is approximately a factor of 3 difference at the latest times between the scenarios with PBH DM and without. With M4, we do not see a sizable difference in the kinetic gas temperature history given the presence of PBHs, but we can identify percent level differences in these the M4 and no-PBH cosmology at the latest times. 

\begin{figure}
    \centering
    \includegraphics[width=\linewidth]{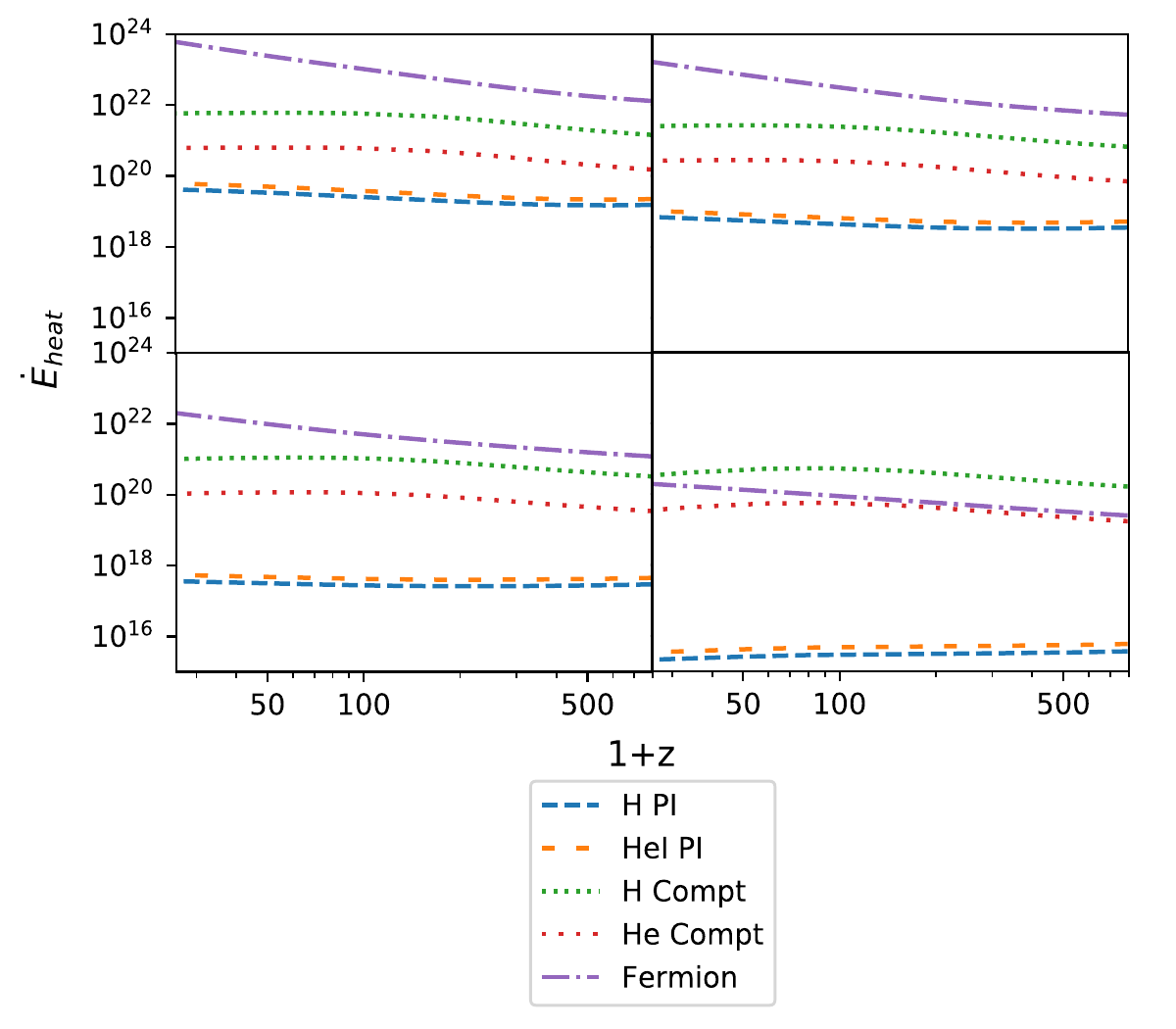}
    \caption{Rates of energy injection [in eV/s] from different heating channels (M1: top left, M2: top right, M3: bottom left, M4: bottom right). Here we see a wide spread in orders of magnitudes of heating rates. With M1, M2 and M3, we see the fermion contributions are most substantial, with rates near $10^{23}$ -$10^{22}$ eV/s that increase with redshift. The Compton scattering contribution with the Hydrogen in the IGM is then next important with rates of $10^{21}$ eV/s, and Helium Compton scattering contributes a few percent to the overall Compton scattering heating rates.}
    \label{fig:EdotHeat}
\end{figure}

To dive deeper into the driving mechanism behind this extreme heating for lighter masses and trivial effects for the heavier mass, we show the different energy injection rates from each of the heating channels in Figure~\ref{fig:EdotHeat}. Here, each panel corresponds to a different PBH mass. In the all but the heaviest mass, we see the heating rates are dominated by processes starting with the fermion Hawking radiation spectrum. As we increase mass, all channels get suppressed, leading to the lessening temperature effects. The decrease in the fermion heating effects can be largely attributed to the inverse Compton cooling effects discussed in Section~\ref{sec: interactions}, as well as the overall dampening of the Hawking radiation spectrum. 

In addition, we find the Compton scattering effects to be the second leading contribution to IGM heating. This is a contrast to our previous intuition; even though there is a low energy tail in the Hawking radiation photon spectrum correction that dominates most energies, we find the higher energy primary spectrum to be the largest contributing factor. Interestingly, we see that accounting not only Hydrogen but also Helium scattering processes are integral to creating a complete picture of IGM heating from PBH DM in this mass range. 

\section{21 cm signal}
\label{sec: 21cm}

\subsection{Implementation}
Another potential avenue for PBHs to modify the IGM history is the 21 cm line evolution. This signal corresponds to the spin-flip transition of hydrogen in the 1s state, and the excitation temperature is referred to as the spin temperature. The potential influences of the spin temperature are the CMB, atomic collisions, and Lyman$\alpha$ scattering coefficients (see \cite{21cosmology} for a review of 21 cm cosmology).

The spin temperature is set by a balance of radiative transitions in the hyperfine line; optical pumping by Lyma-$\alpha$ photons; and collisional coupling:
\begin{equation}
    T_{s}^{-1} = \frac{T_{CMB}^{-1} + x_a T_{a}^{-1} + x_c T_{K}^{-1}}{1 + x_a + x_c},
    \label{eq:Ts}
\end{equation}
where $T_{a}$ is the temperature from the Lyman-$\alpha$ radiation field (which is approximately equal to $T_{K}$ for our interests), $x_{a}$ is the dimensionless coupling due to Lyman-$\alpha$ scattering, and $x_c$ is the dimensionless coupling due to atomic collisions. 

The standard expression for $x_c$ is given by 
\begin{equation}
    x_c = \frac{T_{*}}{A_{10}T_{CMB}} \left(\kappa_{1-0}^{HH}(T_k)n_{\rm H} + \kappa_{1-0}^{eH}(T_k)n_{e} + \kappa_{1-0}^{pH}(T_k)n_{p}\right)
\end{equation}
(see Ref.~\cite{21cosmology}, Eq.~10), where the $\kappa$s are the spin-relaxing collision rate coefficients for different species (H+H, $e^-$+H, and $p^+$+H, respectively).

During the Dark Ages --- after recombination, but prior to astrophysical sources of ultraviolet radiation --- there are no sources of Lyman-$\alpha$ radiation.
Therefore, in this era, we set the Lyman-$\alpha$ coupling to $x_\alpha=0$ for the ``without PBHs'' scenario.

The Lyman-$\alpha$ coupling with PBHs could be nontrivial, so we proceed by calculating could also be influenced by the existence of PBHs by using the equation
\begin{equation}
    x_{\alpha} = S_{\alpha} \frac{J_{\alpha}}{J_{C}},
\end{equation}
where $J_\alpha$ is the specific intensity of Lyman-$\alpha$ photons (originating from hydrogen atom excitations that ultimately trace to Hawking radiation);
\begin{equation}
J_{C} = 1.12 \times 10^{-10} \left(\frac{1+z}{20}\right)\, {\rm cm}^{-2}\,{\rm s}^{-1}\,{\rm sr}^{-1}{\rm Hz}^{-1}
\end{equation}
is a flux normalization; and the dimensionless parameter $S_\alpha$ describing the shape of the Lyman-$\alpha$ line \cite{2004ApJ...602....1C, 2006MNRAS.367..259H} is given by 
\begin{equation}
    S_{\alpha} = e^{-0.37 \sqrt{1+z} T_{k}^{-2/3}} \left(1+ \frac{0.4}{T_{k}} \right)^{-1},
\end{equation}
(with $T_K$ in K) again from Ref.~\cite{21cosmology}.

In order to calculate $J_\alpha$ from PBHs, we must calculate the number of Lyman-$\alpha$ excitations produced either directly from the PBH, or through resultant radiation from PBH Hawking radiation interacting in the IGM. 

The direct Lyman-$\alpha$ photons production rate from the black holes are simple to calculate:
\begin{equation}
    \dot N_{\rm Ly \alpha,  prim} = \omega \frac{dN_{\gamma}(\omega_{\rm Ly\alpha})}{d\omega \,dt} ,
    \label{eq:lya0}
\end{equation}
which is a constant for each PBH mass, determined by the inner bremsstrahlung $\mathcal{O}(\alpha_{\rm EM})$ Hawking radiation spectra. 

To compute the rate of Lyman-$\alpha$ photons produced from secondary interactions, we need to account for excitations produced as a byproduct of initial Hawking radiation photons either photoionizing or Compton scattering, and those produced by initial Hawking radiation fermions exciting the IGM. 

The Lyman-$\alpha$ excitation rate from secondary processes starting with a Hawking radiation photon is given by 
\begin{eqnarray}
    \dot N_{ly \alpha, \rm sec,\gamma} &=& \int \frac{dN_{\gamma}}{d\omega\, dt} (1-e^{-\tau(z)}) p_{\rm ex} f_{\rm Ly\alpha}\nonumber \\
    && \times \sum_{i=H, He} g_{i,\rm PI} (\hbar \omega - E_{0,i})\frac{\hbar \omega-E_{0,i}}{E_{\rm ex}} + g(H,{\rm compt})\frac{h_{IC}(\hbar\omega - E_{o,H})}{E_{\rm ex}}\,d\omega,
    \label{eq:lyap}
\end{eqnarray}
where $E_{ex}$ is the excitation energy of Hydrogen, and $p_{ex}$ is the probability of excitations given in \cite{ShullVanSteenberg} and $f_{\rm Ly\alpha}=0.416$ is the Lyman-$\alpha$ oscillator strength. (We use this as a proxy for the fraction of excitations that go to 2p and hence produce Lyman-$\alpha$ photons. Energy loss from high-energy particles as described by the Bethe-Bloch equation are distributed among transitions according to the oscillator strengths with slight corrections due to the variation of logarithmic factors. At IGM densities, excitations to 3p will almost all result in H$\alpha$ + 2-photon emission, which does not contribute to $J_\alpha$ \cite{2006MNRAS.367..259H, 2006MNRAS.367.1057P}.)

The Lyman-$\alpha$ excitation rate starting with the Hawking-radiated fermions is given by 
\begin{equation}
    \dot N_{\rm Ly \alpha, sec,e^{\pm}} = \int \frac{dN_{e^{\pm}}}{dh\, dt} \frac{h_{IC}(h-m_ec^2)}{E_{ex}}p_{ex}f_{\rm Ly\alpha}\,dh.
    \label{eq:lyae}
\end{equation}

We then use Eqs.~(\ref{eq:lya0}), (\ref{eq:lyap}) and (\ref{eq:lyae}) to calculate the specific flux from PBHs given by
\begin{equation}
    J_{\alpha} = \frac{(1+z)^3 \rho_{DM}c }{m_{DM}4\pi \omega_{ly\alpha}H(z)} (\dot{N}_{ly\alpha, \rm prim} + \dot{N}_{Ly-\alpha, \rm sec, \gamma} + \dot{N}_{Ly-\alpha, \rm sec, e^{\pm}}).
\end{equation}
With these ingredients, it becomes possible to assess the effect of PBHs on the 21 cm spin temperature signal given in Eq.~(\ref{eq:Ts}). 

Typically the observable of interest for the 21 cm signal is given by the differential brightness from the CMB: 
\begin{equation}
    \delta T_b = \frac{T_s - T_{CMB}}{1+z}(1-e^{-\tau}) .
\end{equation}

\subsection{Results}

\begin{figure}
    \centering
    \includegraphics[width=0.7\linewidth]{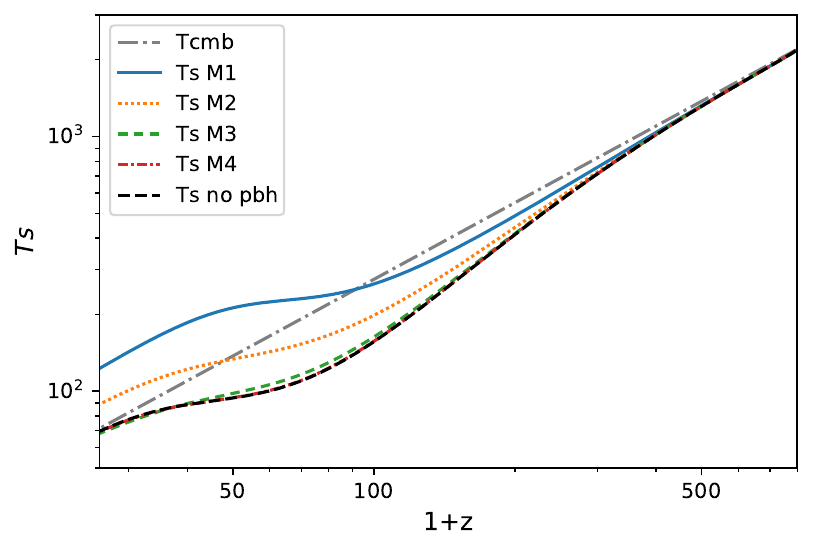}
    \caption{Spin Temperatures [in K] with all PBH masses (again M1 is blue, M2 orange, M3 green, and M4 red solid lines), no PBHs (in black dashed line), and the reference CMB temperature (in gray dashed-dotted line).  We again see both temperatures surpassing the CMB around the same redshifts as the kinetic gas temperature. Here we more minimal deviations to the spin temperature with PBHs as compared to the no-PBH cosmology than with the ionization fraction and kinetic gas temperature, though we see factors of 3 differences at maximum for M1, and sub-percent level differences in the heaviest mass case.}
    \label{fig:Ts}
\end{figure}

\begin{figure}
    \centering
    \includegraphics[width=\linewidth]{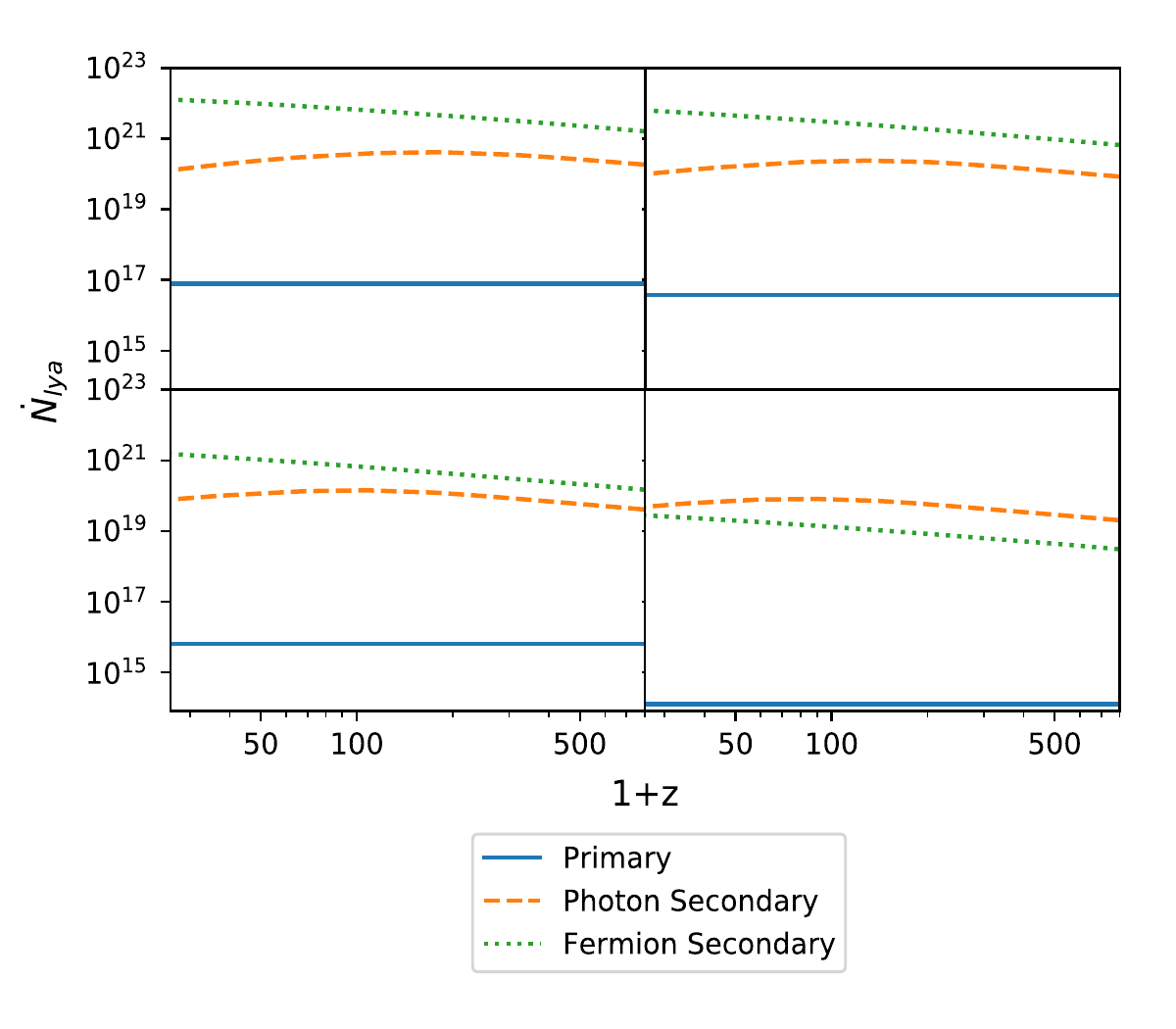}
    \caption{This figure shows the Lyman-$\alpha$ production rates of the different mass PBHs (M1: top left, M2: top right, M3: bottom left, M4: bottom right) split by creation channel (primary Lyman-$\alpha$ in blue, secondary processes starting with photon spectrum in orange, and secondary processes starting with the fermion spectrum in green). These plots show the same trends to other channel plots in this paper: highest overall rates for lowest masses, fermion contributions dominating for most masses, non-direct photon effects having percent-level changes to the over all rates, and trivial contributions from primary photons.}
    \label{fig:LyaChannels}
\end{figure}

\begin{figure}
    \centering
    \includegraphics[width=0.8\linewidth]{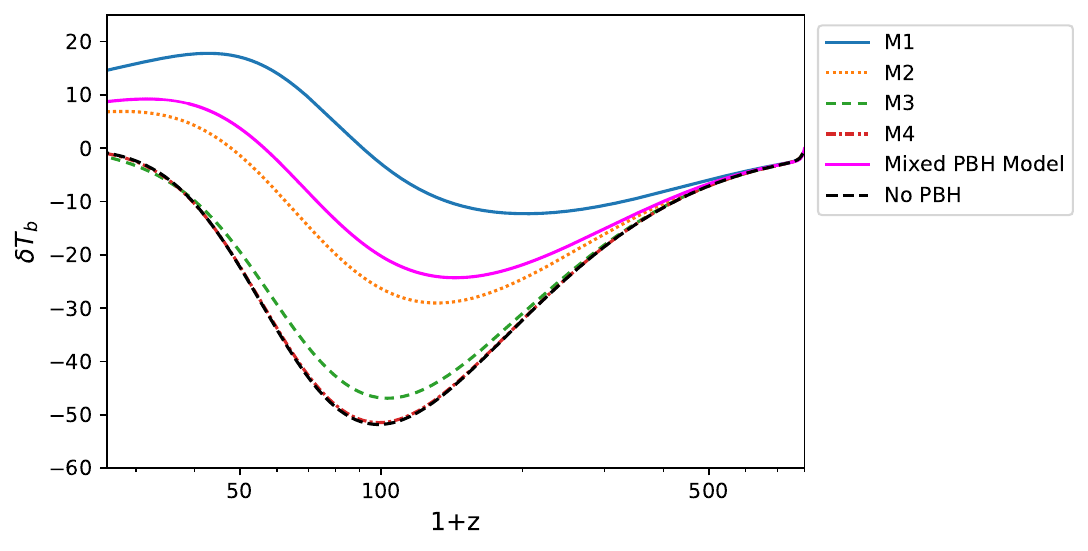}
    \caption{Here we demonstrate the differential brightness temperature [in mK] from the CMB for all masses (M1 solid blue, M2 dotted orange, M3 dashed green, M4 dot-dashed red) and compare this to a non-PBH scenario (in tightly-dashed black). We see the largest two masses have minimal deviations from the non-PBH cosmology. There are deviations from the non-pbh cosmology scenario for the lightest two masses where we see an overall positive signal at late times. In addition to the single-mass PBH distributions shown in previous plots, we also demonstrate in magenta a mixed PBH mass model which is outlined in the discussion. }
    \label{fig:Tb}   
\end{figure}

We show in Figure~\ref{fig:Ts} the spin temperature for our range of PBH masses, with the CMB temperature and an example spin temperature without PBHs as a reference. Again, we see the motif that the lowest PBH masses have the largest impact on our results; the spin temperature can vary by factors of 3 from the no-PBH cosmology case. Most interestingly, M1 has a spin temperature that exceeds the CMB, which can be difficult to achieve in other cosmological scenarios. The largest two PBH masses has much less impact on the spin temperature, but can still reach percent level changes from the no PBH cosmology. 

Again, to try to better understand the temperature results we show the different Lyman-$\alpha$ excitation channels, which we show in Figure~\ref{fig:LyaChannels}. Here the HR fermions once again is responsible for the largest impact in most masses, with the secondary photon HR processes also displaying a meaningful contribution. We also show the suppression of the fermion contribution as we increase the mass, again due to the HR spectral suppression, and the IC Cooling crossover energy calculation. 

Finally, we show in Figure~\ref{fig:Tb} the differential brightness temperature $\delta T_{\rm b}$, our parameter of interest for 21 cm observations. Because this signal is extremely sensitive to gas temperatures, we see evidence for deviations from non-PBH DM cosmologies in all masses. The most pronounced is in M1, where we see deviations as early as $z\sim 500$. We also see the positive values of the differential brightness --- indicates the temperature surpasses that of the CMB --- with no negative values after the zero-crossing. This again would be an extremely interesting cosmological probe if we could measure it observationally; this would be an indicator of some form of an early  efficient heating or excitation mechanism which we show PBHs are capable of in this study. At its peak, M1 shows differential brightness temperatures of +17mK, with maximal discrepancies of 53mK between the M1 PBH and no PBH cosmologies.  

As we consider larger PBH masses, our redshift window for differential brightness differences decreases, but there is no PBH DM mass in this study where PBHs have no difference on this differential brightness temperature in any redshift region; we see varations of 0.5 mK for M4 The sub-percent level deviation between the highest mass and non-PBH dark matter does show that it could be difficult to use the differential brightness signature as a constraint for higher mass PBH dark matter with future observations.

\section{Discussion}
\label{sec: discussion}

\begin{figure}
    \centering
    \includegraphics[width=\linewidth]{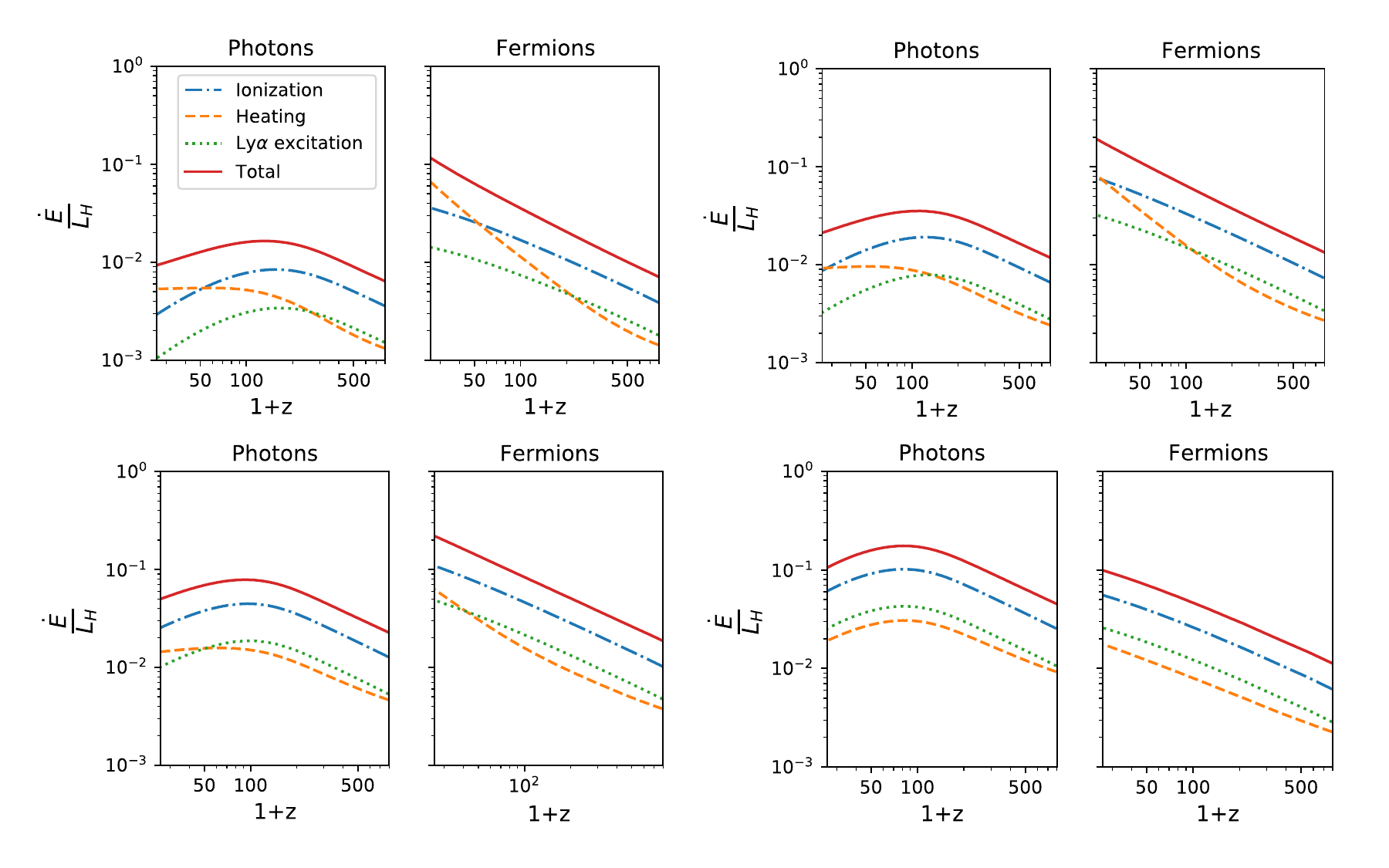}
    \caption{Energy Budget of Black hole that goes into ionizing (blue dot-dashed line), heating (orange dashed line) and exciting (green dotted) the IGM. Results split by mass (M1: top left, M2: top right, M3: bottom left, M4: bottom right), and whether the processes begin with the photon or fermion HR spectrum (left or right subplot). Here we see that we track process that account for anywhere from a 0.1 to 10s of percents of the photon and fermion luminosities. As we move to later times, the processes we track for the fermions become more efficient, whereas the photons have a turnover point due to the optical depth increase over time as the universe stretches. The ionization effects tend to be the most efficient processes, though latest times show the IGM heating to surpass the ionization energy budget.} 
    \label{fig:Energy Budgetting}
\end{figure}

This work investigated different mechanisms through which PBH dark matter can leave imprints on the IGM history through modifying the ionization fraction, increasing the kinetic gas temperature, and creating signatures in the 21 cm differential brightness temperature. One important note about the black holes in this mass range is that they are incredibly small, with Schwarzschild radii of $\mathcal{O}(10^{-14} - 10^{-13})$ meters. Contrary to most astrophysical black holes, these small PBHs are not effective at accreting matter, which we can demonstrate by considering the Bondi accretion radius, 
\begin{equation}
     r_{\rm Bondi} \sim \frac{GM}{c_{\rm sound} ^2} \sim r_{\rm Schwarzchild} \frac{m_{\rm proton} c^2}{kT_k}.
\end{equation}
For concreteness, we can show the Bondi radius of our smallest mass PBH. Using the gas temperature results we obtained with our results discussed section \ref{sec: thermal}, the IGM will reach temperatures of about $T_K \sim 250$~K around $z= 75$, implying a Bondi radius of about $1.5\times10^{-3}$\, m. 
If we consider the hydrogen density to be $\mathcal{O}(10^{-5})$ m$^{-3}$ around $z=$75, that means the number of gas particles within a Bondi radius is $\sim 0.001$: thus instead of ``fluid'' accretion, any accretion process would be dominated by occasional binary collisions between individual atoms and the PBH. 
Therefore, we can treat Hawking radiated fermions as directly being injected into the IGM, rather than passing through a cloud of accreting material. 

This study aimed to give a detailed modeling of how Hawking radiation from PBH dark matter would impact the history of the IGM through ionization, gas temperature, or Lyman-$\alpha$ excitations. The results here do have some limitations. While we assume all dark matter is in the form of PBHs, current constraints do not allow $f_{PBH}$=1 at the lowest masses. These bounds would need to be modified to directly relate these results to a consistent cosmological framework. However, previous bounds use Hawking radiation modeling inconsistent with the treatment in this work and in \cite{HR1, HR2}, so there is potential for these bounds to be modified in the future.
Nonetheless, we can use the observable results in this paper as an upper limit for quantifying PBH effects on the IGM from a specific mass, some of which have several orders of magnitude differences with and without PBHs. This study assumes all PBHs are formed with a specific mass, but PBH mass distributions from more physically motivated formation mechanisms often have extended mass ranges where any particular mass only describes a fraction of the dark matter (see \cite{2024Univ...10..444H} for a review of PBH formation). As such, even if current PBH dark matter abundance bounds exclude these low mass PBHs from being the entirety of dark matter, even a small fraction of dark matter existing in this low asteroid-mass range could give nontrivial observables through the IGM history features we describe in this study. To demonstrate this, in figure \ref{fig:Tb} 

As a reference for an extended mass result, we can assume a uniform distribution over the log of the mass ($df/dM \propto M^{-1}$) for the PBH masses considered in this work. The differential brightness signal from this mixed PBH mass model shows a diminished overall effect relative to the signal from the M1 model, but this extended model shows larger deviations from the no-PBH cosmology than all other single mass distributions. This provides compelling evidence that even small amounts of low-mass PBHs could produce detectable 21cm observables.

An additional outcome of this study is the ability to map the energy emitted budget from a black hole to its impacts on the IGM as a function of time, shown in Figure~\ref{fig:Energy Budgetting}. The different energy rates for different processes are compared luminosity of the PBHs in photons and charged fermions ($e^\pm$). The story here is slightly different than previous results. We see here at the lightest black hole masses, less of the total photon luminosity budget gets propagated into processes we consider in this study, with a maximum of few percent contribution. In the highest mass, we see an order of magnitude increase in the percent contribution from the photon luminosity to the ionization, heating, and excitation processes. In the charged fermion cases, we see at early times, most of the luminosity budget is not in these processes, but as we approach later times, IGM heating, ionization, and excitations can account for over 10 percent of the charged fermion Hawking luminosity.

One limitation of this study --- and opportunity for further study --- is the uniform treatment of dark matter densities, and assume uniformity in our gas densities. This is appropriate for mean quantities at early times where the density perturbations are small. An extension of this study could include a more dynamic treatment of densities, both of dark matter and of the IGM. This modification could prove useful for drawing conclusions on PBH dark matter constraints using current or future 21 cm surveys such as EDGES \cite{edges}, SARAS 3 \cite{Saras}, AMEGO \cite{AMEGO}, and others yet to come. 
\section*{Acknowledgments}

Sincerest thanks to to Rob Claassen, Makana Silva, Gabriel Vasquez, Cara Nel, Bowen Chen, Nihar Dalal, and Maura Kelley for their insightful discussions on this work. C.H.\ and E.K.\ received supported by the David \& Lucile Packard Foundation award 2021-72096 and NASA grant 22-ROMAN11-0011. This material is based upon work supported by the U.S. Department of Energy, Office of Science, Office of Workforce Development for Teachers and Scientists, Office of Science Graduate Student Research (SCGSR) program for E.K. The SCGSR program is administered by the Oak Ridge Institute for Science and Education for the DOE under contract number DE‐SC0014664. This manuscript has been co-authored by FermiForward Discovery Group, LLC under Contract No.\ 89243024CSC000002 with the U.S. Department of Energy, Office of Science, Office of High Energy Physics.

\appendix
\section{Appendix}
\subsection{Hawking radiation modeling}
\label{ap:HR}

Here we discuss the specifics of the numerical model for the Hawking radiation spectra. This includes both the ``free-field'' or $\mathcal{O}(1)$ contribution, and some of the leading $\mathcal O(\alpha_{\rm EM})$ corrections (to the photon spectrum, from dissipative processes).

We break the $\mathcal{O}(1)$ photon spectrum from the PBH into three regions: the low energy spectrum which is fit by a 6th order polynomial,  the high energy exponential decay region, and the middle energy peak in the spectrum which is modeled by a gaussian distribution. Specifically, we use
\begin{equation}
    \frac{dN_{\gamma,0}}{d\omega dt} = \frac{1}{\hbar \omega}\frac{dE_{\gamma,0}}{d\omega dt}
\end{equation}
and fit 
\begin{equation}
    \frac{dE_{\gamma,0}}{d\omega dt} = \begin{cases}
      a_0(\hbar \omega)^6 +a_1(\hbar \omega)^5+a_2(\hbar \omega)^4  \\ ~~~~~~~~~~~~~+a_3(\hbar \omega)^3+a_4(\hbar \omega)^2+a_5(\hbar \omega)+ a_6 & \text{if $\hbar \omega \leq 2 T_{H}/k_{b}$}\\
      b_1 e^{-\frac{(\hbar\omega -b_0)^2}{2\sigma^2} } & \text{if $\hbar \omega > 2T_{H}/k_{b}$ and  $\hbar \omega \leq 8T_{H}/k_{b}$}\\
      c_1 e^{-c_0 \hbar\omega} & \text{otherwise},
    \end{cases}    
    \label{eq:EGT}
\end{equation}
where all coefficients are numerically fitted to primary data and shown in Table~\ref{tab:phot}.

\begin{table}
    \centering
    \begin{tabular}{|c|c|c|}
    \hline
        Energy Range & Parameter & Fitted value \\
        $\hbar\omega/k_{\rm b}T_{\rm H}$ & & \\
        \hline
        \multirow7*{low ($<2$)} &
         $a_0$& $2.17113727\times 10^{-7}$  \\
        & $a_1$& $-2.76745993\times 10^{-6}$ \\
        & $a_2$& $1.52587228\times 10^{-5}$ \\
        & $a_3$& $-9.09411563\times 10^{-6}$ \\
        & $a_4$& $3.74442373\times 10^{-6}$ \\
        & $a_5$& $-9.32934317\times 10^{-7}$ \\
        & $a_6$& $1.04729029\times 10^{-7}$ \\ \hline
         \multirow3*{medium (2--8)} &
         $b_0$& $6.08905731$ \\
        & $b_1$& $3.46186597\times 10^{-3}$ \\
        & $\sigma$ &  $1.57138321$ \\ \hline
         \multirow2*{high ($>8$)} &
         $c_0$ & $-0.75064976$ \\
        & $\ln c_1$ & $-0.31713716$ 
         \\
         \hline
    \end{tabular}
    \caption{Primary photon Hawking radiation energy spectrum in units of the Hawking temperature; see Eq.~(\ref{eq:EGT}).}
    \label{tab:phot}
\end{table}

The first order correction to the Hawking radiation photon spectrum is given as 
\begin{equation}
    \frac{dN_{\gamma,1}}{d\omega dt} = d_0*(\hbar \omega)^2 + d_1(\hbar \omega) +d_2 +  \frac{d_3}{(\hbar \omega)},
    \label{eq:EQD}
\end{equation}
where $d$s are fit to the spectrum data and given in table \ref{tab:Oalphatable}. This $(\hbar\omega) ^{-1}$ scaling allows for X-ray production from PBHs, which motivated this study. 

\begin{table}
    \centering
    \begin{tabular}{c|c|c|c|c}
    \hline
        Mass & $d_0$&$d_1$ &$d_2$ &$d_3$  \\
        \hline
        M1 & $-4.37893943\times 10^{-7}$ & $6.52326628\times 10^{-6}$ & $-4.21409345\times 10^{-5}$ & $1.07131885\times 10^{-4}$  \\
        M2 & $-1.40920574\times 10^{-6}$ & $9.87709677\times 10^{-6}$ & $-3.05705632\times 10^{-5}$ &  $5.01611320\times 10^{-5}$ \\
        M3 & $-1.27657366\times 10^{-6}$ & $4.40789198\times 10^{-6}$ & $-7.60569697\times 10^{-6}$ & $8.68675461\times 10^{-6}$ \\
        M4 & $-1.34830288\times 10^{-7}$ & $2.19368169\times 10^{-7}$ & $-8.13672857\times 10^{-8}$ & $1.69749776\times 10^{-7}$ \\
        \hline
    \end{tabular}
    \caption{Model parameters for the ${\cal O}(\alpha)$ corrections to the Hawking radiation photon spectrum, Eq.~(\ref{eq:EQD}). All energies are given in terms of the Hawking temperature. }
    \label{tab:Oalphatable}
\end{table}

We also include the primary fermion Hawking radiation spectrum for the purposes of this work. 

The electron spectrum is given by interpolating 
\begin{equation}
    \frac{dN_{e,0}}{dh dt} = \sum_{k=-10,\, k\neq 0}^{10}\frac{2 k |T_{\frac12, k,h}|^2 }{2 \pi e^{8\pi Mh + 1}}
\end{equation}
between points, where $k = \pm (j+\frac12)$ is an integer encapsulating the angular momentum $j$ and parity $\pm$, $T_{\frac12, k,h}$ is explicitly evaluated in \cite{HR2}, and we fit an exponential tail for $h>12T_{H}$.

\subsection{Interaction equations}
\label{ap:interactions}

We take the reference formula for photoionization cross sections given by \citet{PIcrosssec}:
\begin{equation}
    \sigma_{PI}(\nu) = \sigma_0 \left[\left(x-1\right)^2 + y_w^2 \right]y^{0.05P - 5.5}\left(1+\sqrt{y/y_a}\right)^{-P}
\end{equation}
where $x = \hbar\nu/E_0 -y_0$, $y=\sqrt{x^2 + y_1^2}$, and other parameters defined in Table~\ref{tab:helium photoionization params}. 

\begin{table}
    \centering
    \begin{tabular}{|c | c| c |}
    \hline
      & H &  HeI  \\
     \hline 
     $E_{0}$ [in ]eV] & 0.4298&13.61   \\
     $\sigma_0$ [cm$^2$] & 5.475$\times 10^{-14}$ &9.492$\times 10^{-16}$ \\
     $y_a$&32.88 & 1.469  \\
     $P$&2.963 & 3.188 \\
     $y_w$&0 & 2.039 \\
     $y_0$&0 & 0.4434 \\
     $y_1$& 0& 2.136 \\ \hline
    \end{tabular}
    \caption{Fitting parameters for H and HeI photoionization, as given by \citet{PIcrosssec}.}
    \label{tab:helium photoionization params}
\end{table}

The Klein-Nishina equation for Compton scattering \cite{1929ZPhy...52..853K} is given by
\begin{equation}
    \sigma_{compton} (k)= 2\pi r_{e}^{2} Z \left[ \frac{1+k}{k^2}\left(\frac{2(1+k)}{1+2k} - \frac{\ln(1+2k)}{k} \right) + \frac{\ln(1+2k)}{2k} - \frac{1+3k}{(1+2k)^2}\right],
\end{equation}
where $k\equiv \hbar\omega/m_ec^2$ is the dimensionless photon energy (relative to the electron rest mass energy), and $Z$ is the number of electrons in an atom. This is technically valid only for photon wavelengths small compared to the Bohr radius (otherwise there are bound-state corrections), but the photoionization channel is dominant in cases where this consideration would matter (see \cite{NIST} for cross section comparisons). 

\bibliography{main.bib}

\end{document}